\documentclass[artice,aps,pra,showpacs,twocolumn,superscriptaddress]{revtex4}
\usepackage{graphicx,color}
\usepackage{amsmath,amsthm,amsfonts,amssymb,bm}
\usepackage{times}
\usepackage{epsf}
\usepackage[colorlinks={true}]{hyperref}
\usepackage[T1]{fontenc}
\usepackage[utf8]{inputenc}
\hypersetup{citecolor={blue}, filecolor={blue}, linkcolor={blue}, urlcolor={blue}}

\newcommand{\commentold}[1]{}
\DeclareMathSymbol{:}{\mathpunct}{operators}{"3A}
\usepackage[utf8]{inputenc}
\usepackage[english]{babel}
\usepackage{amsthm}
\theoremstyle{definition}

\begin{document}

\title{Controlling the quantum speed limit time for unital maps via
filtering operations}
\author{S. Haseli}
\email{soroush.haseli@uut.ac.ir}
\affiliation{Faculty of Physics, Urmia University of Technology, Urmia, Iran
}


\date{\today}

\begin{abstract}
The minimum time a system needs to change from an initial state to a final orthogonal state is called quantum speed limit time. Quantum speed limit time can be used to quantify the speed of the quantum evolution. The speed of the quantum evolution will increase, if the quantum speed limit time decreases. In this work we will use relative purity based bound for quantum speed limit time. It is applicable for any arbitrary initial state. Here, we investigate the effects of filtering operation on quantum speed limit time. It will be observed that for some intervals of filtering operation parameter the quantum speed limit time is decreased by increasing filtering operation parameter and for some other intervals it is decreased by decreasing filtering operation parameter.
\end{abstract}

\maketitle
\section{Introduction}\label{Sec1}
The dynamical speed  of a quantum evolution is one of the most important concepts in quantum theory. It also has wide applications in quantum theory, such as quantum communication \cite{Duan,Bekenstein}, quantum metrology \cite{Giovannetti1,Giovannetti2}, optimal
control \cite{Caneva} and etc. Quantum theory sets a bound on speed of the evolution of quantum systems. The minimal time it takes for the quantum system to transform from an initial state to a target state  is known as quantum speed limit (QSL) time.  QSL time determines the maximum speed of dynamical evolution. There have been many attempts to introduce a comprehensive bound for QSL time. In Ref.\cite{Mandelstam}, Mandelstam and Tamm have introduced a bound for QSL time in closed quantum systems, which reads
\begin{equation}\label{MT}
\tau \geq \tau_{Q S L}=\frac{\pi \hbar}{2 \Delta E},
\end{equation} 
where $\Delta E=\sqrt{\langle\hat{H}^{2}\rangle-\langle\hat{H}\rangle^{2}}$ is the variation of energy of the initial state and $\hat{H}$ is time-independent Hamiltonian describing the dynamics of quantum
system. This bound is called MT bound. In Ref.\cite{Margolus}, Margolus and Levitin have introduced
the bound for closed quantum system based on the mean energy $E=\langle\hat{H}\rangle$ as 
\begin{equation}\label{ML}
\tau \geq \tau_{Q S L}=\frac{\pi \hbar}{2 E},
\end{equation}
this bound is called ML bound. Using these two bounds one can obtain a comprehensive bound as follows\cite{Giovannetti3}
\begin{equation}
\tau \geq \tau_{Q S L}=\max \left\{\frac{\pi \hbar}{2 \Delta E}, \frac{\pi \hbar}{2 E}\right\}
\end{equation}
In Refs.\cite{Jones,Deffner,Pfeifer,Pfeifer1} ,the generalizations of the MT and ML bounds to nonorthogonal
states and to driven systems have been determines. The QSL time for the dynamics of open
systems is also investigated in Refs. \cite{Taddei,delCampo,Deffner1}. An unified bound of QSL time including both MT and ML types for non-
Markovian dynamics has formulated in Ref. \cite{Deffner1}. However this bound is used for initial pure state and it is not feasible for mixed initial states. In Ref.\cite{zhang}, the authors introduce the bound for QSL time which can be used for arbitrary initial states.  They obtained a QSL time for mixed initial states by introducing relative purity as the distance measure, which can define  the speed of evolution starting from an arbitrary initial state in the dynamics of open quantum systems.  

QSL time is inversely related to the speed of the quantum evolution. It means that when QSL time is shortened, the speed of the quantum evolution will increase. Too much efforts have been done to obtain a short QSL time\cite{Min,Zhang1,Song,Wu}. It has been shown that memory effects in non-Markovian dynamics of open quantum systems can reduce the QSL time \cite{Xu}. It has been also shown that the external classical driving can speed up the quantum evolution\cite{Zhang1}. In Ref. \cite{Song}, the authors have used dynamical decoupling pulses to increase the speed of the quantum evolution. 
In this work, we will show how the application of  filtering operation can effect QSL time for the case of unital noises.  Filtering operation is defined by a non-trace-preserving map which can increase the entanglement with some probability. However, it is shown that the filtering operation is a very effective scheme to suppress the decoherence \cite{Li}. We will show that for unital noises the quantum speed limit increases by increasing filtering parameter.   The work is organized as follow. In Sec.\ref{Sec2} give
a brief introduction about the relative purity based QSL time for open quantum systems. The results
and discussion is provided in Sec.\ref{Sec3}. Finally,
the paper is closed with a brief conclusion in Sec.\ref{Sec4}
\section{quantum speed limit time for open quantum systems}\label{Sec2}
The dynamic of an open quantum system can be characterized by the time dependent master equation as 
\begin{equation}
\dot{\rho}_{t}=L_{t}\left(\rho_{t}\right),
\end{equation}
where $\rho_t$  is the state of the open quantum system at time $t$ and $L_t$ is the time-dependent positive
generator. The QSL time is the minimal time it takes for a system to evolve from an initial state $\rho_\tau$ at initial time $\tau$ to a final state $\rho_\tau+\tau_D$ at time $\tau+\tau_D$, where $\tau_D$ is the deriving time. In Ref.\cite{zhang}, the authors have used relative purity to introduced the unified bound for QSL time. They have shown that this QSL time is used for arbitrary initial mixed and pure states. One can obtain the relative purity between initial state $\rho_\tau$ and final state $\rho_\tau+\tau_D$ as
\begin{equation}
f\left(\tau+\tau_{D}\right)=\frac{\operatorname{tr}\left(\rho_{\tau} \rho_{\tau+\tau_{D}}\right)}{\operatorname{tr}\left(\rho_{\tau}^{2}\right)}
\end{equation}
For open quantum system, the ML bound state can be obtain as (See Ref.\cite{zhang} for more details)
\begin{equation}\label{a6}
\tau \geq \frac{\left|f\left(\tau+\tau_{D}\right)-1\right| tr \left(\rho_{\tau}^{2}\right)}{\overline{\sum_{i=1}^{n} \sigma_{i} \rho_{i}}},
\end{equation}
where $\sigma_i$ and $\rho_{i}$ are the singular values of $\mathcal{L}_{t}(\rho_{t})$ and $\rho_{\tau}$, respectively and $\overline{\square}=\frac{1}{\tau_{D}} \int_{\tau}^{\tau + \tau_{D}} \square dt$. The MT bound of QSL-time for open quantum systems can be obtain as
\begin{equation}\label{a7}
\tau \geq \frac{\vert f( \tau + \tau_D ) -1 \vert tr (\rho_{\tau}^{2})}{\overline{ \sqrt{\sum_{i=1}^{n} \sigma_{i}^{2}}}}.
\end{equation}
By combining the results for ML and MT bound, one can arrive at the following general result for QSL time 
\begin{equation}\label{a8}
\tau_{QSL}=\max \lbrace \frac{1}{\overline{ \sum_{i=1}^{n} \sigma_{i} \rho_{i}}}, \frac{1}{\overline{ \sqrt{\sum_{i=1}^{n} \sigma_{i}^{2}}}} \rbrace \times \vert f( \tau + \tau_D ) -1 \vert tr (\rho_{\tau}^{2}).
\end{equation} 
In Ref.\cite{zhang}, the authors have shown that  the ML bound of the QSLT is tighter than MT bound for open quantum systems. From Eq.\ref{a8}, it is obvious that QSL time is always smaller than deriving time $\tau_D$. QSL time is inversely related to speed of the quantum evolution. This means that the speed of quantum evolution increases when the QSL time is shortened, and vice versa.
\section{Control of the QSL time with filtering
operation}\label{Sec3}
In this section we want to study the effects of filtering operation on QSL time. At first we will review the notions of filtering operation and unital noise and then give two example to study the influence of the filtering operation on QSL time of the unital noises.
\subsection{Filtering operation}
In the following analysis, we assume that the local filtering operation is implemented on open quantum system at time $t$. The filtering operation can be written
in the computational basis as
\begin{equation}
F=\left(\begin{array}{cc}
{\sqrt{1-k}} & {0} \\
{0} & {\sqrt{k}}
\end{array}\right)
\end{equation}
where $k$ is the filtering operationparameter  with $0<k<1$. When this
operation is performed on quantum system, the final state can be written as
\begin{equation}
\rho_f(t)=\frac{F \rho_t  F^{\dag}}{tr(F \rho_t  F^{\dag})}.
\end{equation}
\subsection{Unital quantum noise}
Any completely positive trace preserving (CPTP) noise $\Phi$ can be represented in Kraus form as 
\begin{equation}
\Phi\left(\rho\right)=\sum_{k} E_{k}^{\dagger} \rho E_{k}
\end{equation}
where $\rho$ is the initial state of the open quantum system and $E_{k}$'s are Kraus operators with $\sum_{k} E_{k} E_{k}^{\dagger}=I$. A CPTP noise $\Phi$ is unital if and
only if $\sum_{k} E_{k}^{\dagger} E_{k}=I$, i.e.  maps the identity operator
to itself in the same space, $\Phi(I) = I$. Here, we will consider two exaple of unital noises which are : phase damping dynamical model and dephasing model with colored noise
\subsubsection{phase damping dynamical model}
Here we consider a two-level quantum system which interacts with  bosonic
environment. The dynamics of quantum system can be describe by following Hamiltonian
\begin{equation}
H=\frac{\omega_{0}}{2}\sigma_3+\sum_k \omega_k b_{k}^{\dag}b_k + \sigma_3 \sum_k (g_k b_{k}^{\dag}+g_{k}^{*}b_{k}),
\end{equation}
where $\sigma_3$ is the Pauli operator in the z-direction, $\omega_{0}$ represents the two-level system
frequency, $b_k$ and $b_{k}^{\dag}$ are the annihilation creation operators, respectively. $g_k$ is the coupling constant between system and bosonic environment. In this model the dynamics of quantum system
system is characterized by the following time-dependent master equation \cite{Palma}
\begin{equation}
\mathcal{L}(\rho(t))=\frac{\gamma(t)}{2}(\sigma_{3}\rho(t)\sigma_3-\rho(t)),
\end{equation}
where $\gamma(t)$ is the time-dependent dephasing rate. In this model, the off-diagonal elements
of the density matrix of quantum system decay with the decoherence
factor $e^{-\Gamma(t)}$, while the diagonal elements remain unchange. For the case in which the temperature of the environment is zero, $\Gamma(t)$ is given by
\begin{equation}
\Gamma(t)=4 \int d \omega J(\omega)\frac{1-\cos\omega t}{\omega^{2}},
\end{equation}
where $J(\omega)$ is the spectral density of the environment \cite{Palma}. IHere, we consider the Ohmic-like spectral density for the environment 
\begin{equation}
J(\omega)=\omega_{c}^{1-s}\omega^{s}e^{-\frac{\omega}{\omega_c}},
\end{equation}
where $\omega_c$ is the cutoff frequency and $s$ is Ohmicity parameter. Based on the value of Ohmicity parameter, the environment is sub-Ohmic  ($s < 1$), Ohmic ($s = 1$), and
super-Ohmic ($s > 1$).  In this model the dynamic is non-Markovian  when $s \in [2.5,5.5]$ \cite{Haseli1}. The model can be characterized by the following Kraus operators
\begin{equation}
 E_1(t) = \sqrt{\frac{1+e^{-\Gamma(t)}}{2}}\sigma_{0} , \quad E_2(t) = \sqrt{\frac{1-e^{-\Gamma(t)}}{2}}\sigma_{3} . \\
\end{equation}
After applying filtering operation on evolved density matrix, the final density matrix is obtained as 
\begin{equation}
\rho_f(t)=\left(
\begin{array}{cc}
 \frac{1-k}{2} & \frac{1}{2} p_t \sqrt{1-k} \sqrt{k} \\
 \frac{1}{2} p_t \sqrt{1-k} \sqrt{k} & \frac{m}{2} \\
\end{array}
\right),
\end{equation}
where $p_t=e^{-\Gamma(t) }$. From Eq.(\ref{a8}), the QSL time is obtained as 
\begin{equation}
\tau_{QSL}=\sqrt{\frac{k(1-k)}{2}}\frac{p_\tau(p_{\tau+\tau_D}-p_\tau)}{\frac{1}{\tau_D}\int_{\tau}^{\tau+\tau_D}\dot{p}_t dt}.
\end{equation}
\begin{figure}[!]  
\centerline{\includegraphics[scale=0.5]{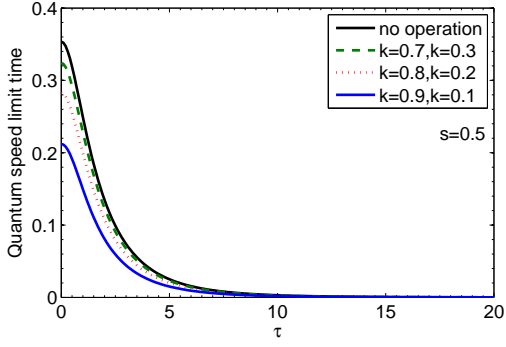}}
\vspace*{8pt}
\caption{Quantum speed limit time as a function of initial time $\tau$ for sub Ohmic environment $s=0.5$ for different value of filtering operation parameter.}\label{Fig1}
\end{figure}
In Fig.(\ref{Fig1}), QSL time is plotted as a function of initial time $\tau$ for sub Ohmic environment $s=0.5$ for different value of filtering operation parameter $k$. It is obvious that  the increasing of the filtering operation parameter in the  region, $0.5 < k < 1$ leads to quantum speedup of quantum evolution. While increasing the filtering operation parameter in the region  $0 < k < 0.5$ slowdown the quantum evolution.  
\begin{figure}[!]  
\centerline{\includegraphics[scale=0.5]{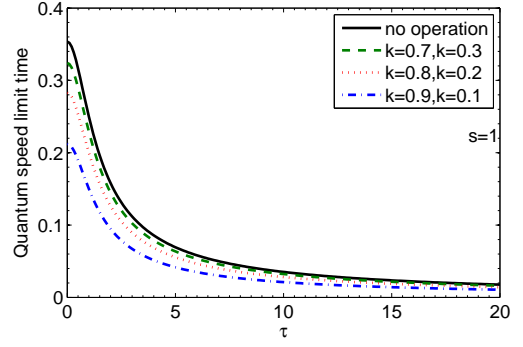}}
\vspace*{8pt}
\caption{Quantum speed limit time as a function of initial time $\tau$ for Ohmic environment $s=1$ for different value of filtering operation parameter. }\label{Fig2}
\end{figure}
Fig.\ref{Fig2}, represents the QSL time for Ohmic environment $s=1$ with different value of filtering operation parameters. As can be seen, the QSL time is decreased by  increasing of the filtering operation parameter in the  region, $0.5 < k < 1$. So, in this region increasing filtering operation parameter speedup the quantum evolution. We  also observe that the QSL time is increased by increasing filtering operation in the region $0 < k < 0.5$. 
\begin{figure}[!]  
\centerline{\includegraphics[scale=0.5]{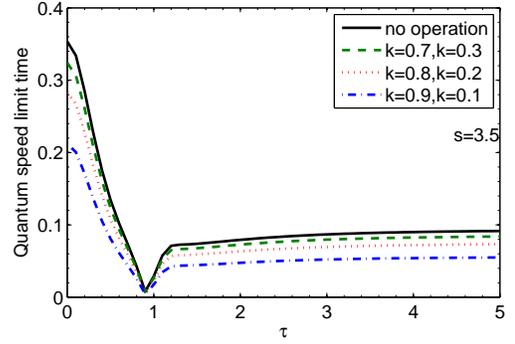}}
\vspace*{8pt}
\caption{Quantum speed limit time as a function of initial time $\tau$ for super Ohmic environment $s=3.5$ for different value of filtering operation parameter. }\label{Fig3}
\end{figure}
Fig.\ref{Fig3}, shows the QSL time for super Ohmic environment $s=3.5$ with various value of filtering operation parameters. As can be seen, the QSL time is decreased by  increasing of the filtering operation parameter in the  region, $0.5 < k < 1$. We also observe that the QSL time is increased by increasing the filtering operation parameter in the region  $0 < k < 0.5$.  The fluctuation observed in the QSL time is due to the non-Markovian property of the noise.
\subsubsection{dephasing model with colored noise}
Now, we consider the interaction between a two-level quantum system with an environment which has the property of a random telegraph signal noise. The dynamics of quantum system is characterized
by time dependent Hamiltonian 
\begin{equation}
H(t)=\sum_{m=1}^{3} \Gamma_m(t) \sigma_m,
\end{equation}
where $\sigma_m$'s are the Pauli  operators in ($x,y,z$) directions. $\Gamma_m(t)$'s are random variable
which follow the statistics of a random telegraph signal. $\Gamma_{m}(t)$ depends on the random variable  $n_{m}(t)$ as $\Gamma_{m} (t)=\alpha_m n_m(t)$, Where $n_{m}(t)$ has a Poisson distribution with  an average value equal to $t/2\tau_m$ and $\alpha_m$'s are coin-flip random variables that  randomly can have values $\pm \alpha_m$. Here, we consider dephasing model with colored noise with $\alpha_1=\alpha_2= 0$ and $\alpha_3=\alpha$. In this model, the dynamics can be defined via the following Kraus operators
\begin{equation}
 E_1(t) = \sqrt{\frac{1+\Lambda_t}{2}}\sigma_{0} , \quad E_2(t) = \sqrt{\frac{1-\Lambda_t}{2}}\sigma_{3}, \\
\end{equation}
where $\Lambda_t=e^{-t/2 \Delta}[\cos(\mu t/2 \Delta)+\sin(\mu t/2 \Delta)/\mu]$, $\mu=\sqrt{(4 \alpha \Delta )^{2}-1}$. The dynamic is non-Markovian for $\alpha \Delta \geq 1/2$. After applying filtering operation on evolved density matrix, the final density matrix is obtained as 
\begin{equation}
\rho_f(t)=\left(
\begin{array}{cc}
 \frac{1-k}{2} & \frac{1}{2} \Lambda_t \sqrt{1-k} \sqrt{k} \\
 \frac{1}{2} \Lambda_t \sqrt{1-k} \sqrt{k} & \frac{k}{2} \\
\end{array}
\right),
\end{equation}
From Eq.(\ref{a8}), the QSL time is obtained as 
\begin{equation}
\tau_{QSL}=\sqrt{\frac{k(1-k)}{2}}\frac{\Lambda_\tau(\Lambda_{\tau+\tau_D}-\Lambda_\tau)}{\frac{1}{\tau_D}\int_{\tau}^{\tau+\tau_D}\dot{\Lambda}_t dt}.
\end{equation}
\begin{figure}[!]  
\centerline{\includegraphics[scale=0.5]{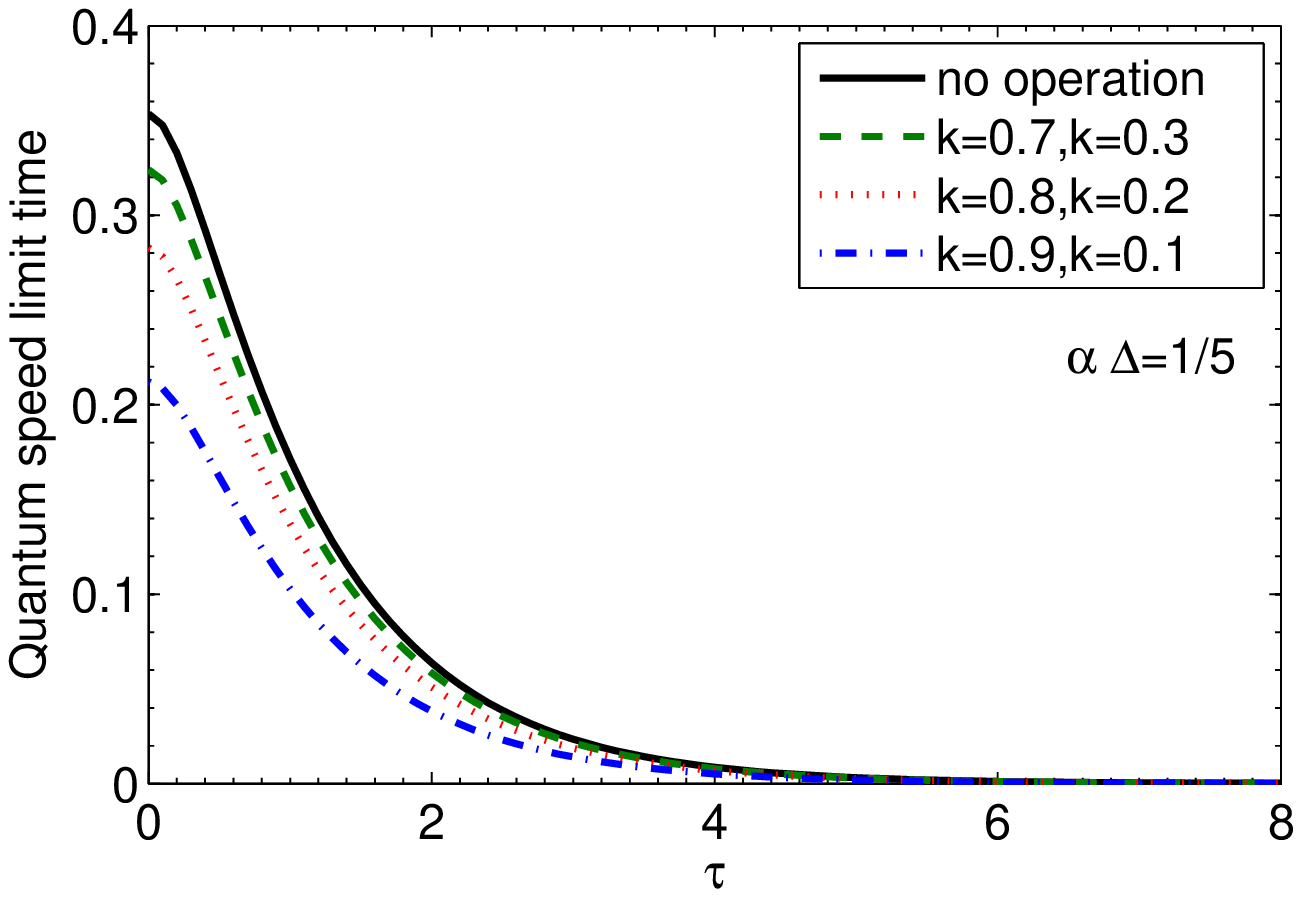}}
\vspace*{8pt}
\caption{Quantum speed limit time as a function of initial time $\tau$ with $\alpha \Delta=1/5$ for different value of filtering operation parameter. }\label{Fig4}
\end{figure}
\begin{figure}[!]  
\centerline{\includegraphics[scale=0.5]{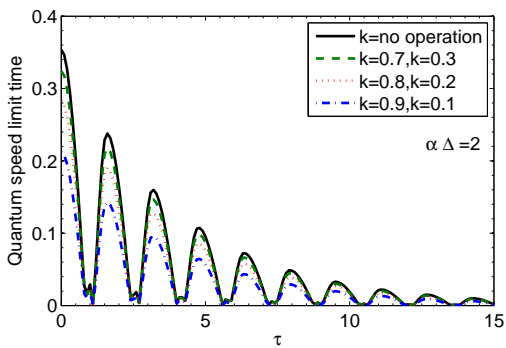}}
\vspace*{8pt}
\caption{Quantum speed limit time as a function of initial time $\tau$ with $\alpha \Delta=2$ for different value of filtering operation parameter. }\label{Fig5}
\end{figure}
In Fig.(\ref{Fig4}), QSL time is plotted as a function of initial time $\tau$ for $\alpha \Delta=1/5$  with different value of filtering operation parameter $k$. It is obvious that  the increasing of the filtering operation parameter in the  region, $0.5 < k < 1$ leads to quantum speedup of quantum evolution. While the increasing of the filtering operation parameter in the  region, $0 < k < 0.5$ leads to slowdown of quantum evolution.    

Fig.\ref{Fig5}, represents the QSL time as a function of initial time for  $\alpha \Delta=2$ with different value of filtering operation parameters. As can be seen, the QSL time is decreased by  increasing of the filtering operation parameter in the  region, $0.5 < k < 1$. While the QSL time is increased by increasing filtering operation parameter in the region, $0< k < 0.5$. The fluctuation observed in the QSL time is due to the non-Markovian property of the noise.
\section{Conclusion}\label{Sec4}
In this work we investigated the effects of filtering operation on QSL time. We observed that the QSL time is decreased by  increasing of the filtering operation parameter in the  region, $0.5 < k < 1$. In other words in this region one can speedup the quantum evolution by increasing the filtering operation parameters. We can also observe that the quantum speed limit time is decreased by decreasing filtering operation parameter  in the  region, $0 < k < 0.5$. In other word, in this region one can slowdown the quantum evolution by increasing the filtering operation parameters.


\end{document}